\documentclass[preprint,12pt]{elsarticle}
\usepackage{amssymb}
\usepackage{amsmath}
\usepackage{url}
\usepackage{xcolor}
\usepackage{float}
\usepackage{subcaption}
\usepackage{booktabs}
\usepackage{textcomp}
\usepackage{setspace}
\usepackage{caption}
\usepackage{array}
\usepackage{longtable}
\usepackage{siunitx}
\usepackage{pslatex}
\usepackage[letterpaper,top=2.5cm,bottom=2.5cm,left=3cm,right=2.5cm,marginparwidth=1.75cm]{geometry}
\usepackage[colorlinks=true, urlcolor=blue, linkcolor=black]{hyperref}

\journal{Tunneling and Underground Space Technology}

\begin{document}

\begin{frontmatter}

\title{Dynamical Modeling of Temperature and Smoke Evolution in a Thermal-Runaway Event of a Large-Format Lithium-ion Battery in a Mine Tunnel}


\author[1]{Khadia Omar Said}
\ead{kos5600@psu.edu}
\author[2]{Yukta Pareek}
\ead{ybp5153@psu.edu}
\author[2]{Satadru Dey}
\ead{skd5685@psu.edu}
\author[1]{Ashish Ranjan Kumar \corref{cor1}}
\ead{awk5528@psu.edu}

\cortext[cor1]{Corresponding author}

\affiliation[1]{organization={Department of Energy and Mineral Engineering, The Pennsylvania State University},
    addressline={58 Pollock Road},
    city={University Park},
    postcode={16801},
    country={United States}}
\affiliation[2]{organization={Department of Mechanical Engineering, The Pennsylvania State University},
    addressline={Reber Building},
    city={University Park},
    postcode={16801},
    country={United States}}


\begin{abstract}
Large-format lithium-ion batteries (LIBs) provide effective energy storage solutions for high-power equipment used in underground mining operations.
They have high Columbic efficiency and minimal heat and emission footprints. 
However, improper use of LIBs, accidents, or other factors may increase the probability of thermal runaway (TR), a rapid combustion reaction that discharges toxic and flammable substances.  
Several such incidents have been documented in underground mines.
Since repeatable TR experiments to uncover the transient-state propagation of TR are expensive and hazardous, high-fidelity models are usually developed to mimic the impact of these events. 
They are resource-intensive and are impractical to develop for many scenarios that could be observed in a mine.
Therefore, dynamic models within a reduced-order framework were constructed to represent the transient-state combustion event. 
Reduced order models (ROMs) reasonably replicate trends in temperature and smoke, showing strong alignment with the ground-truth dataset.
\end{abstract}

\begin{highlights}
\item Thermal runaway in large-format lithium-ion batteries underground is hazardous.
\item High-fidelity numerical models presenting these combustion events are computationally expensive. 
\item Reduce-order models capture the major trends in the transient-state thermo-fluid parameters.
\end{highlights}


\end{frontmatter}


\section{Introduction}
\label{Intro}

Underground mining operations typically use high voltage electrical or diesel engine-driven equipment for material transport. 
Electrical equipment, such as a tethered shuttle car, is commonly used in underground coal mines that use continuous miners to develop workings and production \cite{skousen2021coal}. 
It drags the electric cable, has limited mobility, and could be hazardous to miners who work around them.
Other underground mines use diesel equipment for their flexibility \cite{rawlins2023underground}.
Diesel particulate matter (DPM) released from these engines is a known carcinogen \cite{chang2017review, weitekamp2020systematic}. 
Therefore, the mining industry is evaluating the application of rechargeable (secondary) batteries in several non-coal unit operations, with haulage being the most promising. 
Lithium-ion batteries (LIBs) are the most commonly used secondary batteries. 
They have expanded considerably in their scope of applications. 
This is due to their high energy density, long cycle life, low memory effect, and high Coulombic efficiency
\cite{kim2019lithium}.
However, there have been multiple incidents of violent failure due to a rapid combustion process known as thermal runaway (TR) \cite{huang2021experimental, feng2015characterization}.
TR events lead to fire, sometimes with explosions, and the emission of several flammable hydrocarbons and other toxic gases that can present hazardous conditions for underground mining personnel \cite{wang2019thermal, koch2018comprehensive}. 

A TR event associated with a large-format battery could be catastrophic for exposed personnel. 
Controlled tests to study these events are hazardous, expensive, and unsuitable for repeated trials. 
Therefore, numerical modeling is a viable alternative. Generally, computational fluid dynamics (CFD) models are developed to mimic a TR event in a facility. 
These high-fidelity models require accurate flow volume geometry and boundary conditions to deliver results in terms of spatio-temporal thermo-fluid parameters \cite{hoelle20233d}.
However, CFD models remain computationally expensive despite recent advances in high-performance computing systems.
Therefore, they cannot be deployed rapidly to analyze complex scenarios, such as large-format battery combustion, especially when there are rapid changes in underground environmental conditions. 
In addition, as thermo-fluid parameters like velocity and temperature fluctuate rapidly because of turbulence, using an averaged value often suffices for the implementation of mine safety protocols. 

This work aims to bridge the aforementioned research gap by contributing to the body of knowledge by developing a reduced-order model (ROM) for fast and accurate modeling of key LIB TR products, including temperature and smoke.
ROMs have previously been developed for convective heat transfer \cite{zucatti2020assessment, rambo2007reduced}.
High-fidelity (ground-truth) data sets for this research were developed by simulating TR LIB in a tunnel using the Fire Dynamics Simulator (FDS) developed by the National Institute of Standards and Technology (NIST) \cite{mcgrattan2013fire}.
The MATLAB System Identification Toolbox was used to generate data-driven state-space models for each node under investigation.
Models were developed for the failure of two LIBs, with capacities of 60 Ah and 243 Ah, respectively, in a mine tunnel \cite{liu2021experimental, peng2020new}.
Longitudinal airflow speeds of 1.0 m/s and 2.0 m/s in the tunnel were considered for this investigation.
Two data sets were concatenated for training and one data set was used for testing. 

\begin{figure}[htbp!]
    \centering
    \includegraphics[width=0.95\textwidth]{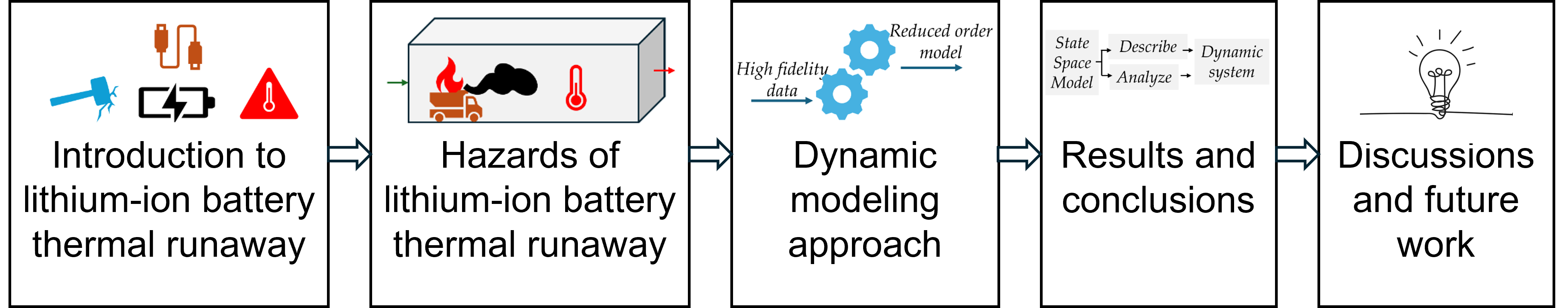}
    \caption{\centering Illustration of the structure of the manuscript showing key focus for each section}
    \label{Structure}
\end{figure}

The paper is structured as follows. Section \ref{Intro} presents an introduction to the dynamic modeling of temperature and smoke from LIB TR. 
Section \ref{Hazards} presents the key hazards emanating from LIB TR, including temperature and smoke, highlighting their potential risks in subterranean facilities. 
Section \ref{Dynamic} elaborates on the method used to develop the dynamic modeling framework. 
Section \ref{Results} describes the results obtained from this work, including FDS simulation, training and testing models, and uncertainty analysis. 
Section \ref{Future} presents the conclusions and avenues for further research as indicated in Fig. \ref{Structure}.

\section{Battery Fire Hazards: Temperature and Smoke Evolution}
\label{Hazards}

\subsection{Environmental impacts of thermal runaway}

LIB TR could be a serious hazard due to rapid heat release and the emission of noxious gases and aerosols, especially in large-format units \cite{somandepalli2014quantification, saada2015causes}.
These events present complex thermochemical reactions with several distinct stages and unique characteristics \cite{feng2014thermal, zhang2023multi}.
Internal short circuits act as the trigger, yet they do not significantly contribute to heat release.
TR could begin at any affected cell located in the pack and spread rapidly to adjacent cells. 
However, the interaction between the anode and cathode in a large-format LIB causes the main exothermic reaction
\cite{feng2019investigating}. 
Temperature differences exceeding 700 \textdegree{C} within a battery have been observed \cite{feng2015characterization}.
The impact of TR on personnel and assets is exacerbated in confined underground mines, where ventilation configurations and geometry critically influence the spread of temperature, gases, and smoke.
Understanding and predicting their propagation and spread promptly is instrumental in the deployment of suppression mechanisms and emergency response planning.

LIB TR also results in the formation of soot, which has been reported to be more toxic to human lungs and neural cells than soot from wood. 
The soot of LiFePO\textsubscript{4} and the ternary battery typically exhibit `core shell' with the former containing elements such as carbon, oxygen, lithium, fluorine, phosphorus and iron, while the latter also contains nickel, cobalt and manganese \cite{xu2024soot}.
A study on LIB TR particulates reported that LIB TR-PM concentrations are approximately six times higher than under ambient conditions. 
A comparison indicated that TR of LIBs of LFP chemistry resulted in a soot concentration of  \( 356.3 \pm 23.2~\mathrm{mg/m^3} \) while NMC had \( 401.1 \pm 30.1~\mathrm{mg/m^3} \) \cite{meister2025evaluating}. 
As such, the timely detection and monitoring of LIB TR together with its products is crucial to deploy mitigation measures and ensure the safety of first responders. 

The temperature profiles during LIB TR have been well documented in the literature.
For example, a 20 Ah LiFePO\textsubscript{4} LIB was subjected to TR by overcharging at 100\% SoC using a constant current and voltage sequence. 
The study reported maximum temperatures of \SI{166.1}{\celsius}, \SI{166.1}{\celsius}, and \SI{139.2}{\celsius} when overcharged at 0.5 C, 0.75 C, and 1 C, respectively \cite{zhang2023multi}.
Another study investigated the TR of 50 Ah and 105 Ah LiFePO\textsubscript{4} batteries. 
Temperature ranges of \SI{150}{\celsius} to \SI{400}{\celsius} and \SI{160}{\celsius} to \SI{540}{\celsius} were reported, respectively, with an increase in heating temperature within the range of \SI{150}{\celsius} to \SI{300}{\celsius} for the two units \cite{zhou2022investigating}.
A research on the stability of various LIB chemistries indicated peak temperatures of \SI{900}{\celsius}, \SI{1000}{\celsius}, and \SI{800}{\celsius} due to local heating of 2.6 Ah LiCoO\textsubscript{2} at 100 \% SoC,  3.4 Ah LiNi\textsubscript{0.8}CO\textsubscript{0.15}Al\textsubscript{0.05}O\textsubscript{2} at 50 \% SoC, and 1.5 Ah  LiFePO\textsubscript{4} at 100\% SoC \cite{kong2021numerical}.

\subsection{Relevant numerical modeling frameworks}
The development of numerical models allows for a parametric study of these combustion events. 
These are usually generated using the finite-volume framework, where the fluid flow volume is discretized into millions of control volumes. 
The propagation of the LIB TR products in the flow volume can be explained using equations of mass, species, momentum, and energy conservation \cite{mcgrattan2013fire}. 
These are shown in Eq.  ~\eqref{eq:mass}--\eqref{eq:energy}.
Here, $\mathbf{\rho}$ is the fluid density ($\mathrm{kg/m^3}$), 
$\mathbf{\vec{u}}$ is the velocity vector ($\mathrm{m/s}$), 
$\mathbf{p}$ is the pressure ($\mathrm{Pa}$), 
${\tau}$ is the viscous stress tensor ($\mathrm{Pa}$), 
$\mathbf{\vec{g}}$ is the gravitational acceleration vector ($\mathrm{m/s^2}$), 
$\mathbf{h}$ is the specific enthalpy ($\mathrm{J/kg}$), 
$\mathbf{k}$ is the thermal conductivity ($\mathrm{W/(m{\cdot}K)}$), 
$\mathbf{T}$ is the temperature ($\mathrm{K}$), and 
$\mathbf{\dot{Q}}$ is the volumetric heat source term due to thermal runaway ($\mathrm{W/m^3}$), 
and $\mathbf{\dfrac{Dp}{Dt}}$ is the material derivative of pressure ($\mathrm{Pa/s}$). 
These are implemented in the FDS software that was used to develop the numerical models for this research.

\begin{subequations}
\begin{equation}
\frac{\partial \rho}{\partial t} + \nabla \cdot (\rho \mathbf{u}) = 0
\label{eq:mass}
\end{equation}

\begin{equation}
\frac{\partial}{\partial t}(\rho Y_l) + \nabla \cdot (\rho Y_l \mathbf{u}) =
\nabla \cdot [(\rho D)_l \nabla Y_l] + \dot{W}''_l
\label{eq:species}
\end{equation}

\begin{equation}
\rho \left( \frac{\partial \mathbf{u}}{\partial t} + (\mathbf{u} \cdot \nabla)\mathbf{u} \right)
+ \nabla p = \rho \mathbf{g} + \mathbf{f} + \nabla \cdot \boldsymbol{\tau}
\label{eq:momentum}
\end{equation}

\begin{equation}
\frac{\partial}{\partial t}(\rho h) + \nabla \cdot (\rho h \mathbf{u})
- \frac{Dp}{Dt} = \dot{q}''' + \nabla \cdot (k \nabla T)
+ \nabla \cdot \sum_l h_l (\rho D)_l \nabla Y_l
\label{eq:energy}
\end{equation}
\end{subequations}

Alternatively, state space models (SSMs) have proven their efficiency in modeling dynamic systems based on internal states, inputs, and outputs using differential or difference models \cite{friedland2012control}. 
These provide a mathematical framework for describing complex systems in terms of differential equations. 
In discrete-time format, SSMs are generally shown as illustrated in Eq. \ref{first} and \ref{second}.
Here $\mathbf{k}$ represents the time index, $\mathbf{x(k)}$ represents the vector of internal states (e.g., temperatures or smoke concentrations at specific locations), $\mathbf{u(k)}$ is the input (e.g., heat release rate), and $\mathbf{y(k)}$ is the output (e.g., measurable temperature or smoke at sensors).
The corresponding matrices are represented by the symbols $\mathbf{A}$-$\mathbf{D}$.
SSMs have been widely used to model various hazards, including road accidents, city fire surveillance, and flooding \cite{fang2021development, gomes2022flood}.
However, despite their proven prowess, limited studies have deployed SSMs to monitor LIB post-TR hazards, especially in underground mines.
This study aims to investigate the efficiency of SSM as an appropriate tool to monitor the evolution of smoke and temperature during LIB TR in a tunnel, as described in the next section. 

\begin{subequations}
\begin{align}
x(k+1) &= A x(k) + B u(k) \label{first} \\
y(k) &= C x(k) + D u(k) \label{second}
\end{align}
\end{subequations}

\begin{figure}[h!]
    \centering    \includegraphics[width=0.6\textwidth]{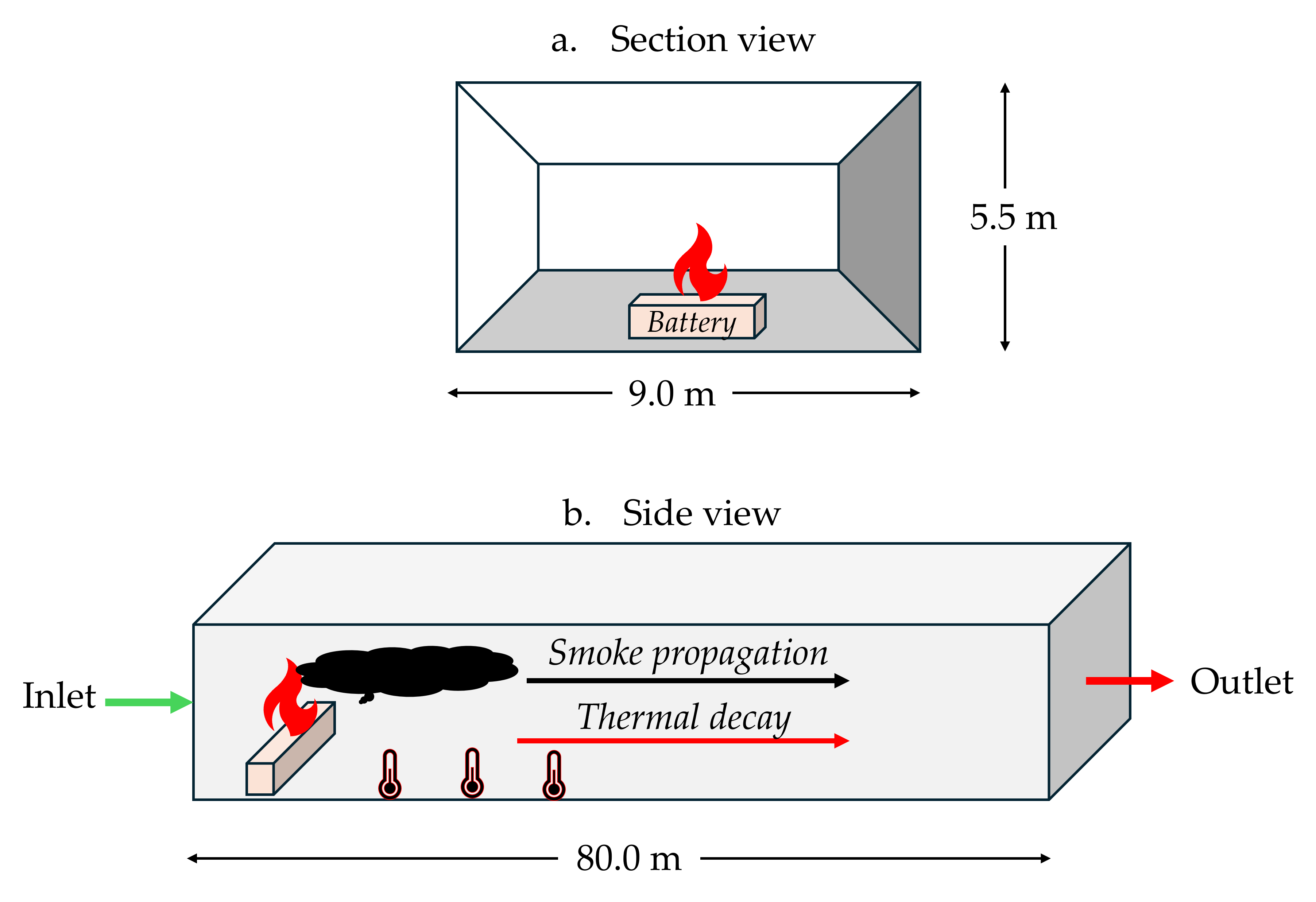}
    \caption{\centering Geometry of the study domain: The first illustration depicts the battery alongside a tunnel cross-section with dimensions of 9.0 m x 5.5 m; the second one shows the direction of airflow with the temperature decaying along it}
    \label{Tunnel_geometry}
\end{figure}



\section{Dynamical Modeling Framework}
\label{Dynamic}

To develop reduced-order models for the dynamic behavior of the smoke and temperature resulting from LIB TR in an underground mine tunnel, numerical simulation and data-driven system identification tools were used.
High-fidelity computational fluid dynamics (CFD) models for two LIB capacities, 60 Ah and 243 Ah, were developed under varying airflow conditions in the tunnel using the Fire Dynamics Simulator software (illustrated in Fig. \ref{Tunnel_geometry}). 
A real-world monitoring scenario was achieved by resolving the spatio-temporal evolution of smoke and temperature with output recorded at strategic nodes downstream of the LIBs.
This data was used as ground truth.

\begin{figure}[h!]
    \centering    \includegraphics[width=0.80\textwidth]{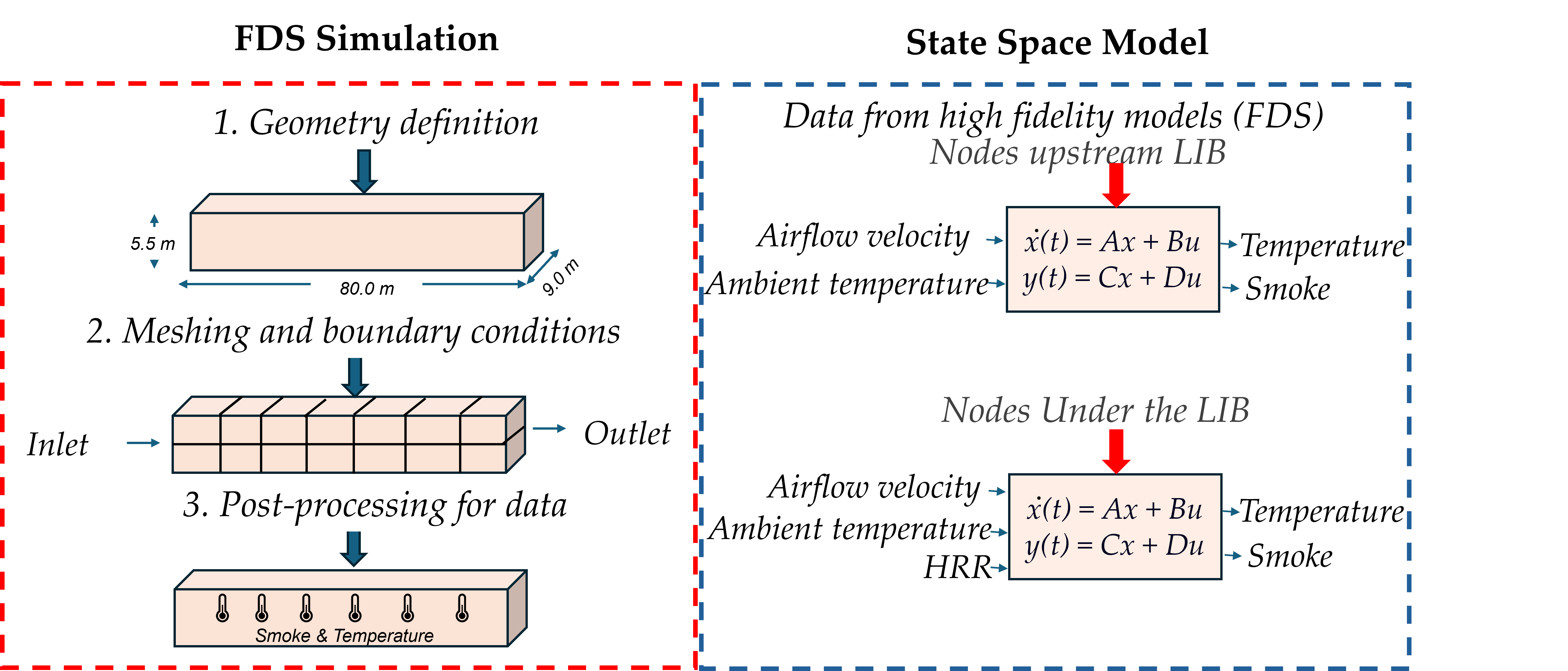} 
    \caption{\centering Summary of employed method: FDS provides the high fidelity ground truth data using the flow volume and boundary conditions to provide the ground-truth data, while SSMs present an efficient pathway to capture the dynamics using the input data and internal states}
    \label{Method}
\end{figure}

A comprehensive data set that encapsulates possible thermal and fluid flow regimes was created through concatenation.
MATLAB's system identification toolbox was used to learn the underlying dynamics from the FDS-generated data to create continuous-state SSMs for both smoke and temperature. 
Two additional data sets were used for model testing.
This enables the assessment of SSM's ability to reproduce the captured dynamic features with a reduced computational complexity. 
A summary of this process is shown in Fig. \ref{Method}.
This section details the procedure used for developing the SSM for temperature and smoke.

\subsection{Data Generation using Fire Dynamics Simulator}
\begin{figure}[h!]
    \centering
\includegraphics[width=0.5\linewidth]{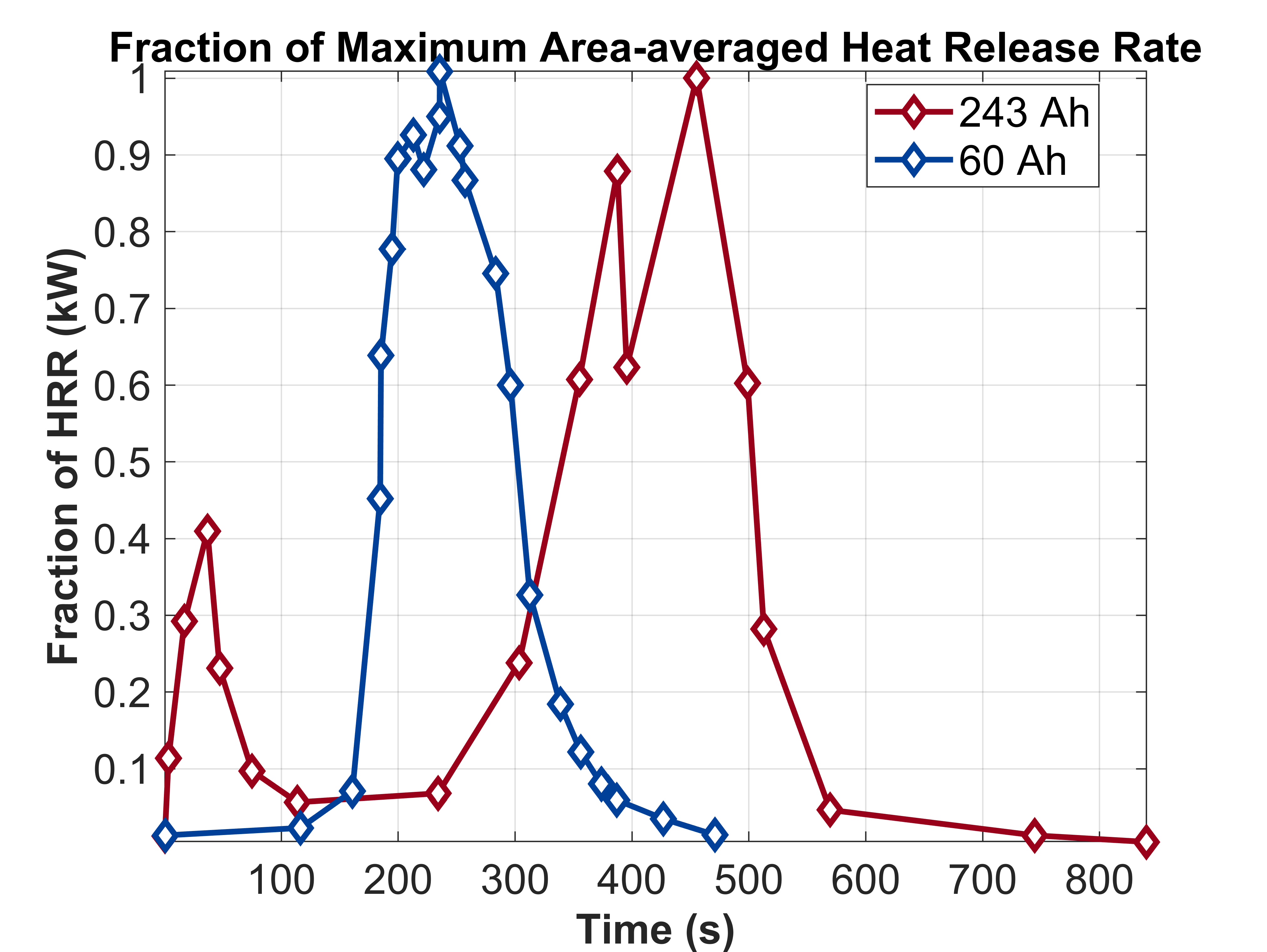}    
    \caption{\centering HRR of the battery capacities under consideration, including 243 Ah and 60 Ah LIBs, presented as a fraction of the maximum area-averaged heat release rates }
    \label{fig:HRR}
\end{figure}

This study simulated the transient-state combustion event presented by LIB TR of two large units appropriate for large equipment (60 Ah and 243 Ah) in an 80.0 m long tunnel,
The tunnel had a cross-sectional area of 9.0 m x 5.5 m as presented in Fig. \ref{Tunnel_geometry}.
The gravitational acceleration was assigned in the vertical direction as -9.81 \( \mathrm{m/s^2} \) to ensure natural entrainment, layering, and buoyancy-driven flows in a real-life setting.
The ambient temperature was set to 15 \textdegree{}C.
The dimensions of the grid cells used to develop the mesh were determined using the length of the fire characteristics (Eq. \ref{Dstar}). 
Here, $\mathbf{\dot{Q}}$ is the peak heat release rate (HRR) (kW), 
$\mathbf{\rho_\infty}$ is the ambient air density (kg/m\(^3\)), 
$\mathbf{C_p}$ is the specific heat capacity (kJ/kg/K), and 
$\mathbf{T_\infty}$ is the ambient temperature (K).
The HRRs for the two battery capacities are presented in Fig. \ref{fig:HRR}.
Once a robust mesh was developed, boundary conditions were assigned. 

\begin{equation}
D^* = \left( \frac{\dot{Q}}{\rho_\infty C_p T_\infty \sqrt{g}} \right)^{2/5}
\label{Dstar}
\end{equation}


The entry and exit of the tunnels were defined using \texttt{\&VENT} to represent the entry and exit boundaries of the simulation domain. 
The LIB, which is the fire source, was modeled as \texttt{\&OBSTRUCTION} at the entrance of the tunnel.
The obstruction was modeled to occupy spatial bounds of \( x=3.0\text{–}6.0~\mathrm{m},\ y=0.0\text{–}10.0,\ \text{and}\ z=0.5\text{–}1.5~\mathrm{m} \)
to represent the geometry of the fire source.
The associated surface conditions were assigned using \texttt{\&SURF ID}, where the combustion surface was assigned the \texttt{\&FIRE SURFACE} properties, while the walls are inert, presented using \texttt{\&INERT} command.
The combustion of LIB TR was assigned the reaction ID of the electrolyte because it has the lowest flashing point and is the most volatile and flammable part of LIBs.
The heat profile was defined using the \texttt{\&RAMP} function, specifying the fractional progression of the peak HRR as a function of the increase in time monotonically.
This ensures that the fire grows over time based on the specified HRR profile, while accurately presenting the convective and radiative behavior of combustion surfaces.
The temperature and smoke probes were placed in the center of the tunnel (that is, 4.5 m from the side walls) and 1.5 m high (average height of the miners) at the entrance and after every 8.0 m, equidistant throughout the tunnel.
The corresponding files stored the boundary conditions.
The simulations were run on a high performance workstation using parallel compute cores. 
Transient temperatures and smoke concentrations for different scenarios were recorded in the corresponding files.

\subsection{State Space Modeling}
Time-series data for LIB TR of two LIB capacities, 60 Ah and 243 Ah, were concatenated to ensure that the model captures a wide range of thermal and smoke dynamics under varying energy densities and failure intensities. 
The concatenated data was then used as training input for system identification in MATLAB for SSM development \cite{matsys}. 
The process used `Grey-Box' system identification, in which physics insights guided input-output parameter selection, and system dynamics were learned from the time-series data.

\begin{figure}[h!]
    \centering    \includegraphics[width=0.6\textwidth]{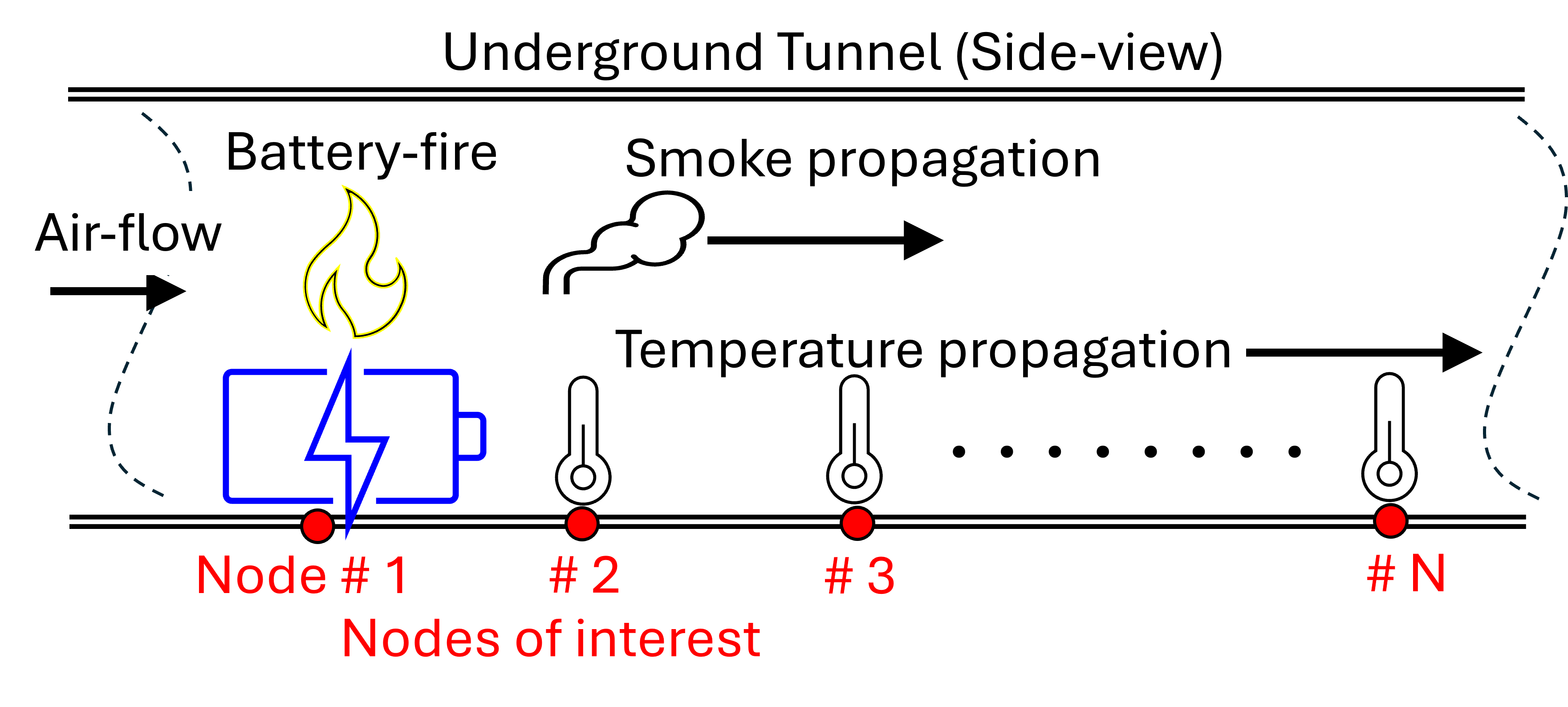}
    \caption{\centering Schematic of the underground mine setup for dynamical modeling purposes. The lithium-ion battery is placed $\sim$ 5.0 m from the mine inlet. Temperature and smoke are monitored at nodes below and downstream of the LIB}
    \label{fig:model}
\end{figure}

The SSM for the node next to the LIB was developed using two key exogenous inputs: (i) HRR, and (ii) ambient temperature. 
These variables were assumed to be the key drivers for localized thermal response, while the ground truth temperature was used as the output.
The downstream nodes were modeled as dependent systems to capture convection and diffusive heat propagation.
For each node $ i $ downstream, the input parameters considered were ambient temperature and temperature output from the preceding node $T(y_{i-1}, t)$ to capture spatial coupling between adjacent nodes (Fig. \ref{fig:model}). 
The output for these nodes was the local node temperature, $T(y_{i}, t)$.
This node-wise cascaded structure allowed modular training and testing of the dynamic model for temperature and smoke.

\section{Results and Conclusions}
\label{Results}
High-fidelity simulation using FDS yielded spatio-temporal temperature and smoke concentration profiles of 60 Ah and 243 Ah LIBs undergoing a full-length thermal runaway.
For these combustion events, a constant airflow rate was maintained in the tunnel.  
As the TR event propagated, the heat and particulate injected in the tunnel followed the HRR curve. 
The evolution of temperature and the transportation of particulates, driven by combined fire-driven buoyancy and forced ventilation, were obtained at the designated nodes under investigation, placed equidistant from the fire source.
The near-field nodes exhibited higher temperature transients, taking shorter durations to reach peak temperature.

\begin{figure}[h!]
    \centering
    \begin{subfigure}[b]{0.475\textwidth}
        \centering        \includegraphics[width=\linewidth]{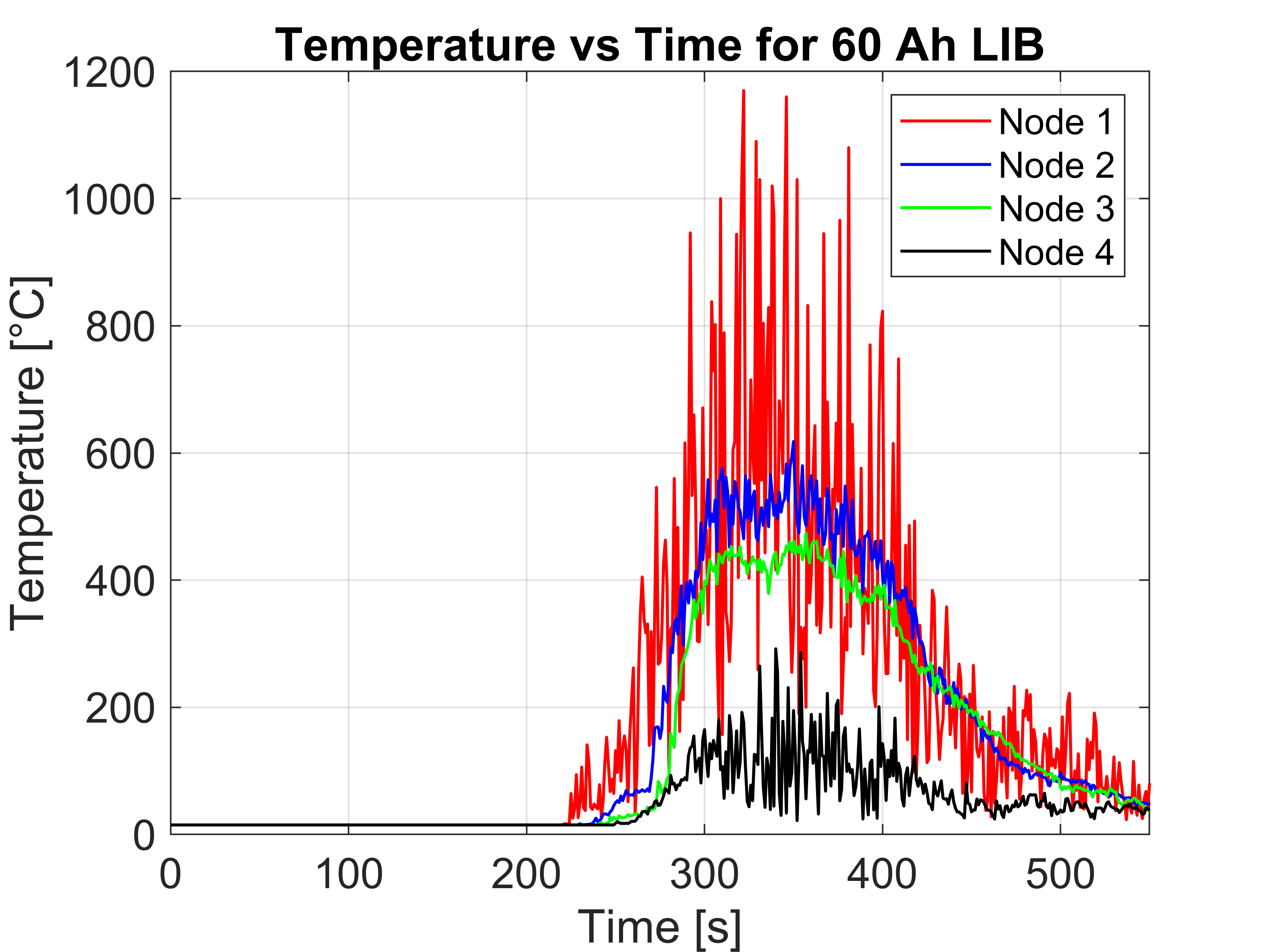}
        \caption{}
        \label{60Ah_T}
    \end{subfigure}
        \hfill
        \begin{subfigure}[b]{0.475\textwidth}
        \centering        \includegraphics[width=\linewidth]{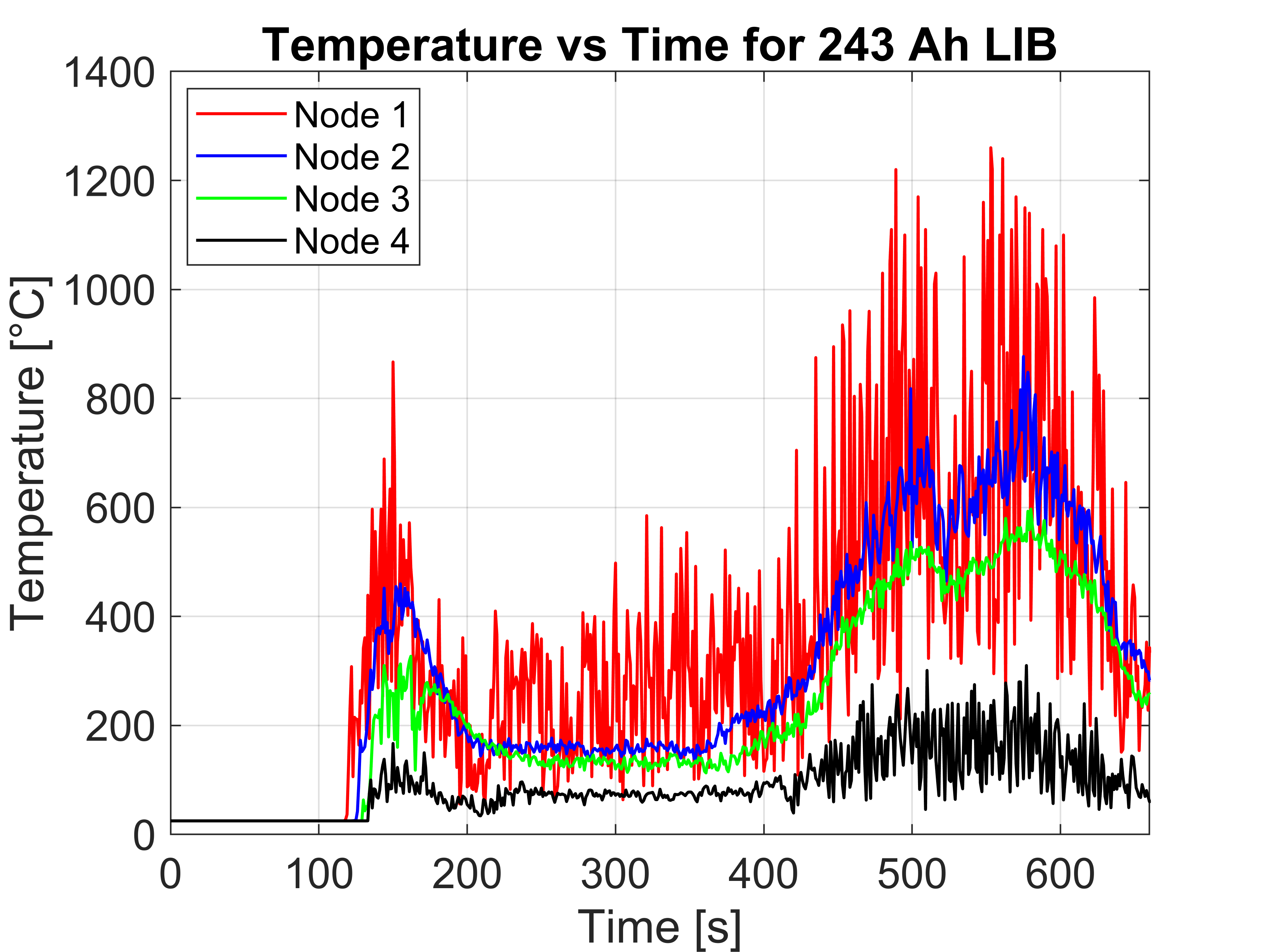}
        \caption{}
        \label{243Ah_T}
    \end{subfigure}
    
    \caption{\centering Temperature plots from airflow velocity of 2.0 m/s from: \textbf{(a)} 60 Ah LIB illustrating increased thermal decay with increased distance from the LIB. Nodes closer to the LIB have higher temperatures than nodes downstream, \textbf{(b)} 243 Ah LIB indicate much higher temperature profiles than 60 Ah. Thermal decay is observed at nodes downstream of the LIB}
    \label{HRR}
\end{figure}

\begin{figure}[t!]
    \centering
    \begin{subfigure}[b]{0.475\textwidth}
        \centering        \includegraphics[width=\linewidth]{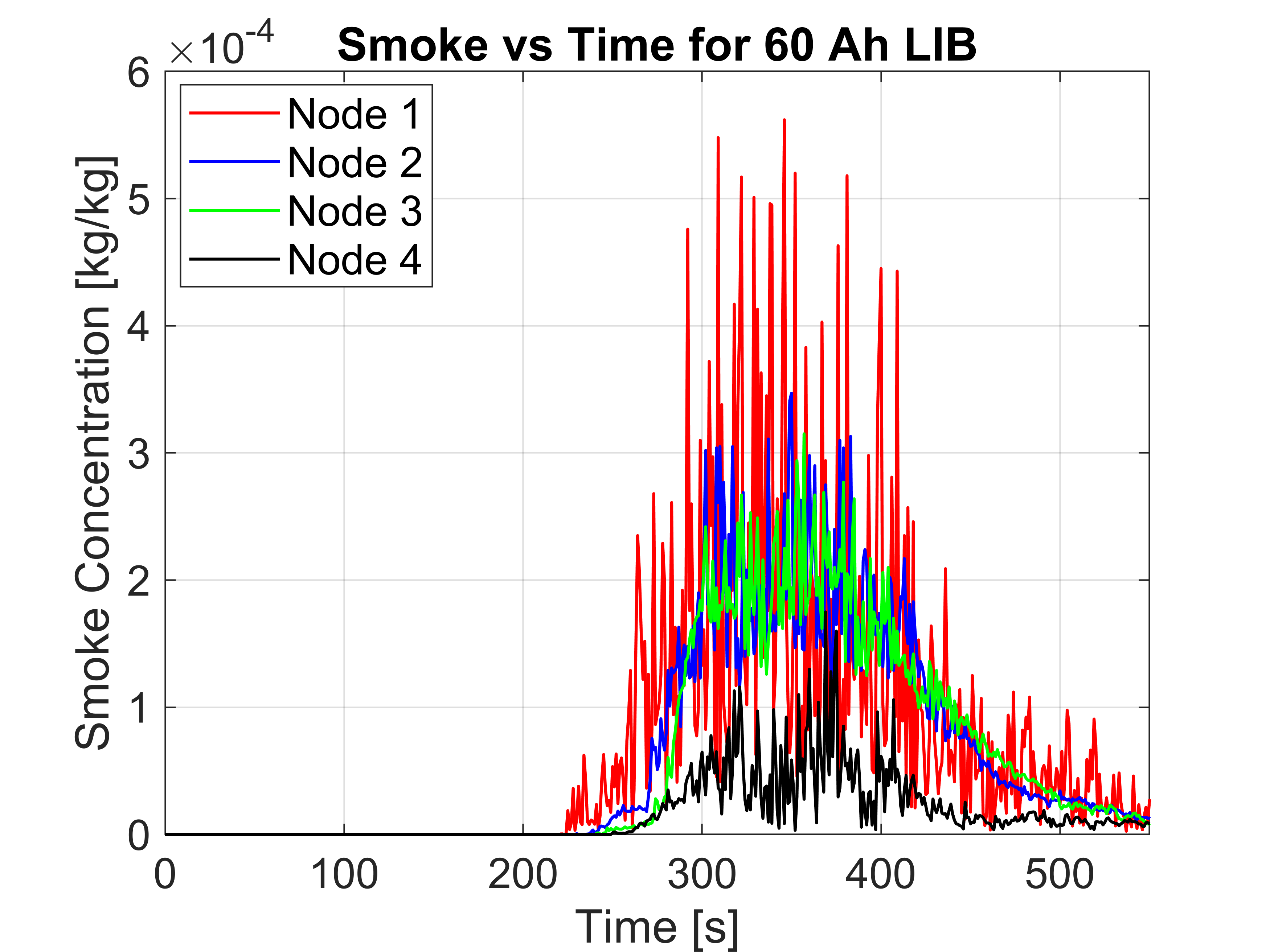}
        \caption{}
        \label{60Ah_S}
    \end{subfigure}
        \hfill
        \begin{subfigure}[b]{0.475\textwidth}
        \centering        \includegraphics[width=\linewidth]{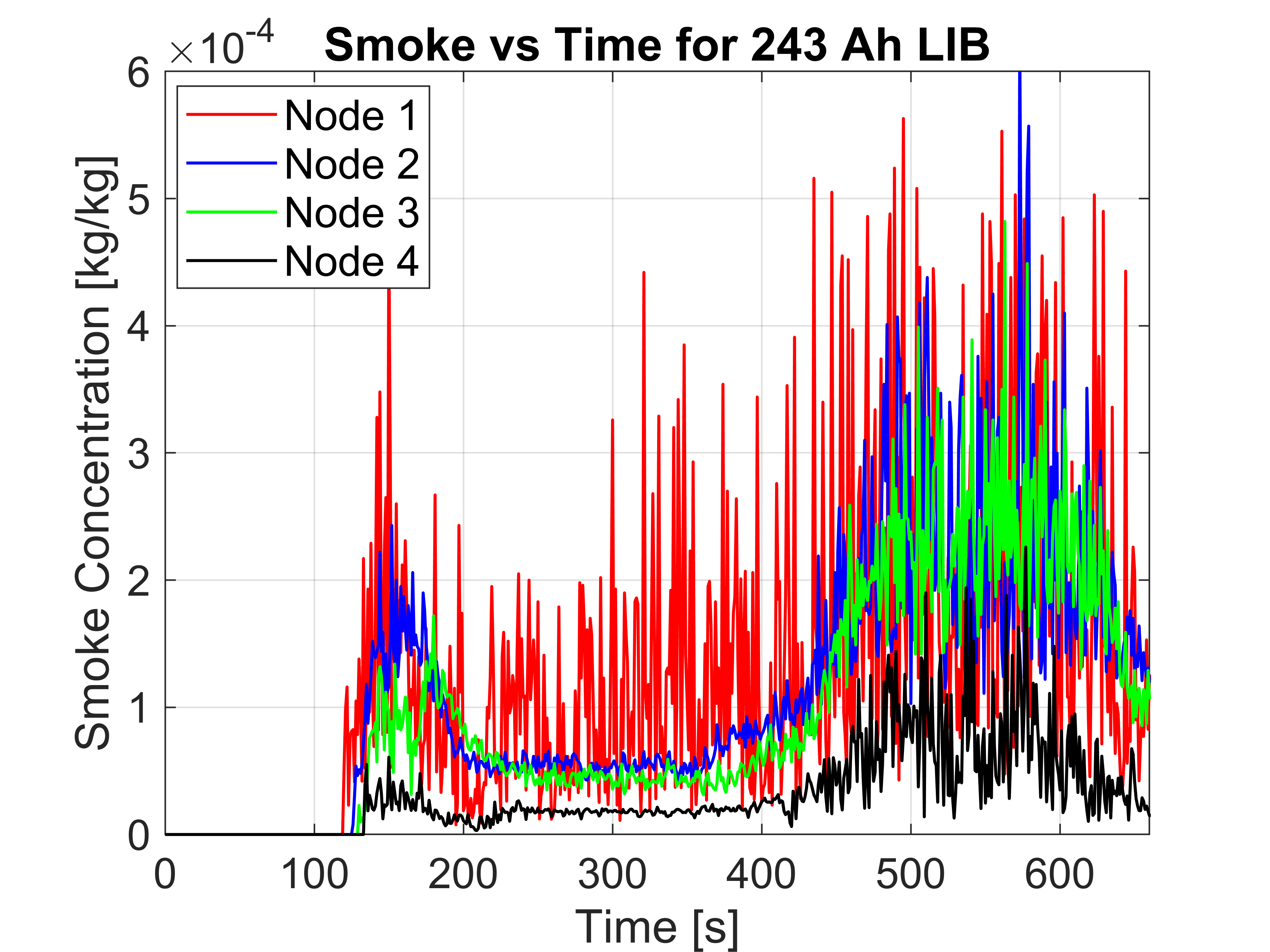}
        \caption{}
        \label{243Ah_S}
    \end{subfigure}
    
    \caption{\centering Smoke plots from airflow velocity of 2.0 m/s from: \textbf{(a)} 60 Ah LIB indicate higher concentration in nodes closer to the LIB than nodes downstream, \textbf{(b)} 243 Ah LIB indicate higher concentrations than 60 Ah}
    \label{HRR}
\end{figure}

For the 60 Ah LIB, the peak temperatures for nodes 1, 2, 3, and 4 were $\sim$ \SI{1180}{\celsius}, \SI{605}{\celsius}, \SI{405}{\celsius}, and \SI{300}{\celsius}, respectively, as indicated in Fig. \ref{60Ah_T}.
In the case of 243 Ah LIB, nodes 1, 2, 3, and 4 exhibited peak temperatures of $\sim$ \SI{1250}{\celsius}, \SI{820}{\celsius}, \SI{600}{\celsius}, and \SI{240}{\celsius}, respectively, as indicated in Fig. \ref{243Ah_T} respectively.
This thermal gradient is caused by the propagation of the thermal plume downstream, which intermixes with ambient air, convective losses to the surrounding area, and attenuation of thermal radiation, which reduces the temperatures further from the source.
Similar trends are observed for smoke concentration, as illustrated in Fig. \ref{60Ah_S} and \ref{243Ah_S}.
The smoke concentration increases as the fuel burns more vigorously. 
As the smoke plume propagates downstream, it becomes diluted, and its concentration progressively decreases. 
This spatial variation reflects the combined effect of convective transport and turbulent mixing in the study domain. 


\subsection{Training and Testing}
\begin{figure}[h!]
    \centering
    \includegraphics[width=0.75\textwidth]{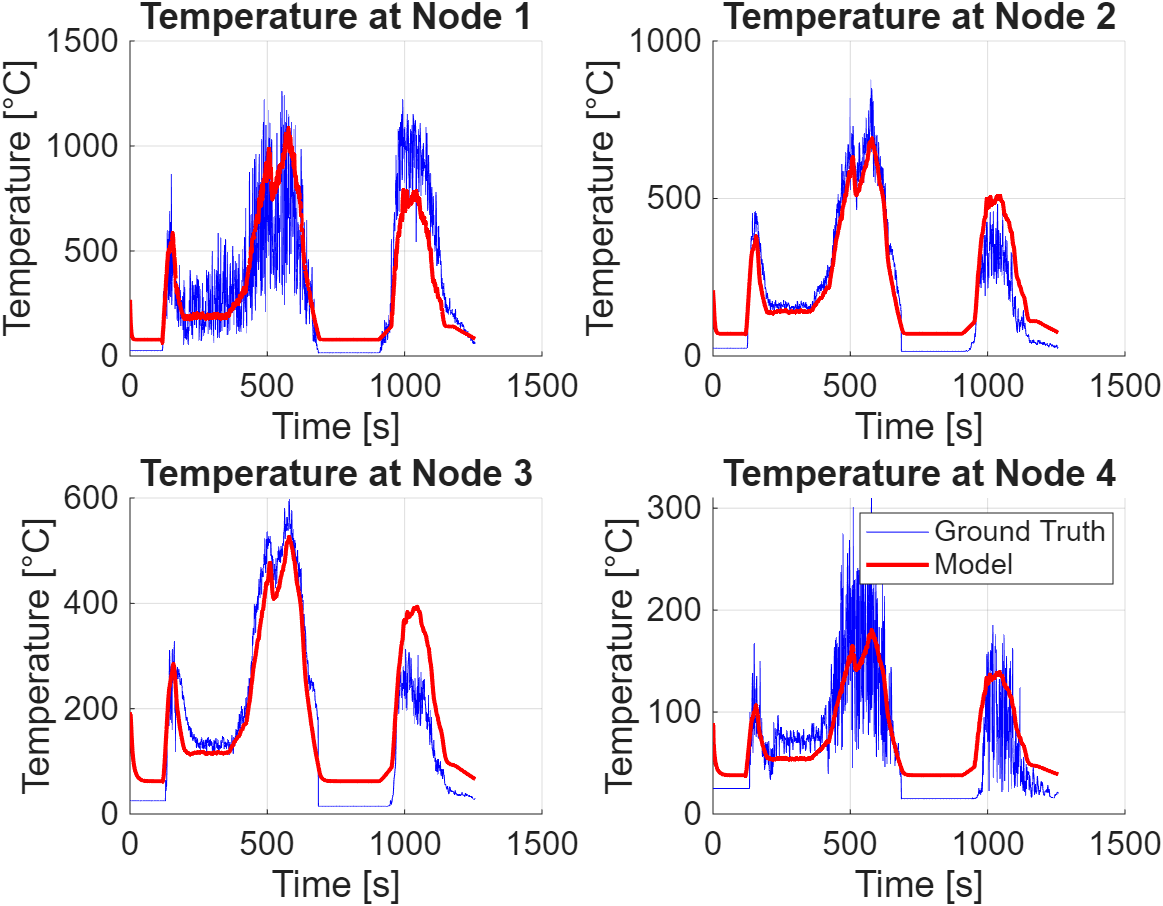}
    \caption{Comparison of the dynamical temperature propagation model and ground truth, for the training dataset}
    \label{Train}
\end{figure}

\begin{figure}[h!]
    \centering
    \includegraphics[width=0.7\textwidth]{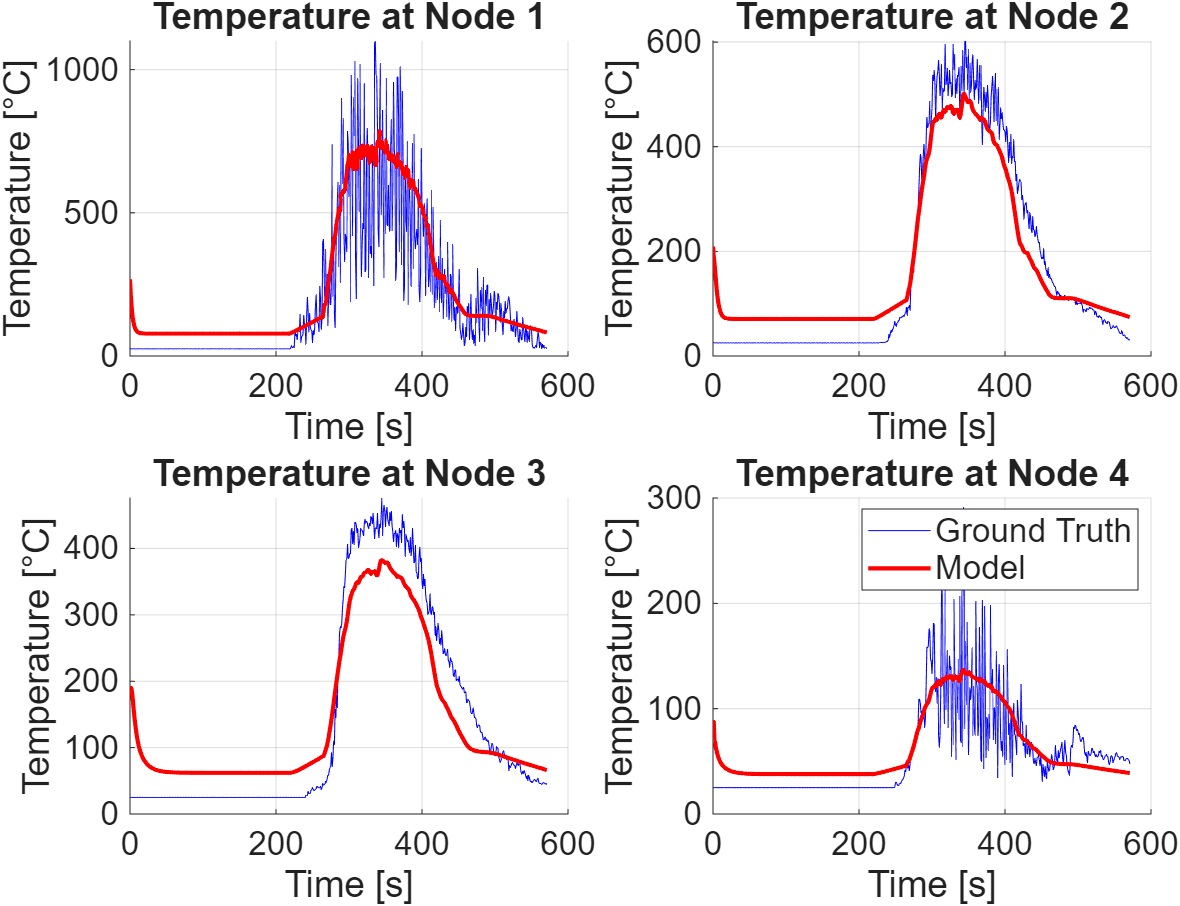} 
    \caption{Comparison of the dynamical temperature propagation model and ground truth for testing dataset}
    \label{Testing}
\end{figure}

\begin{figure}[h!]
    \centering    \includegraphics[width=0.7\textwidth]{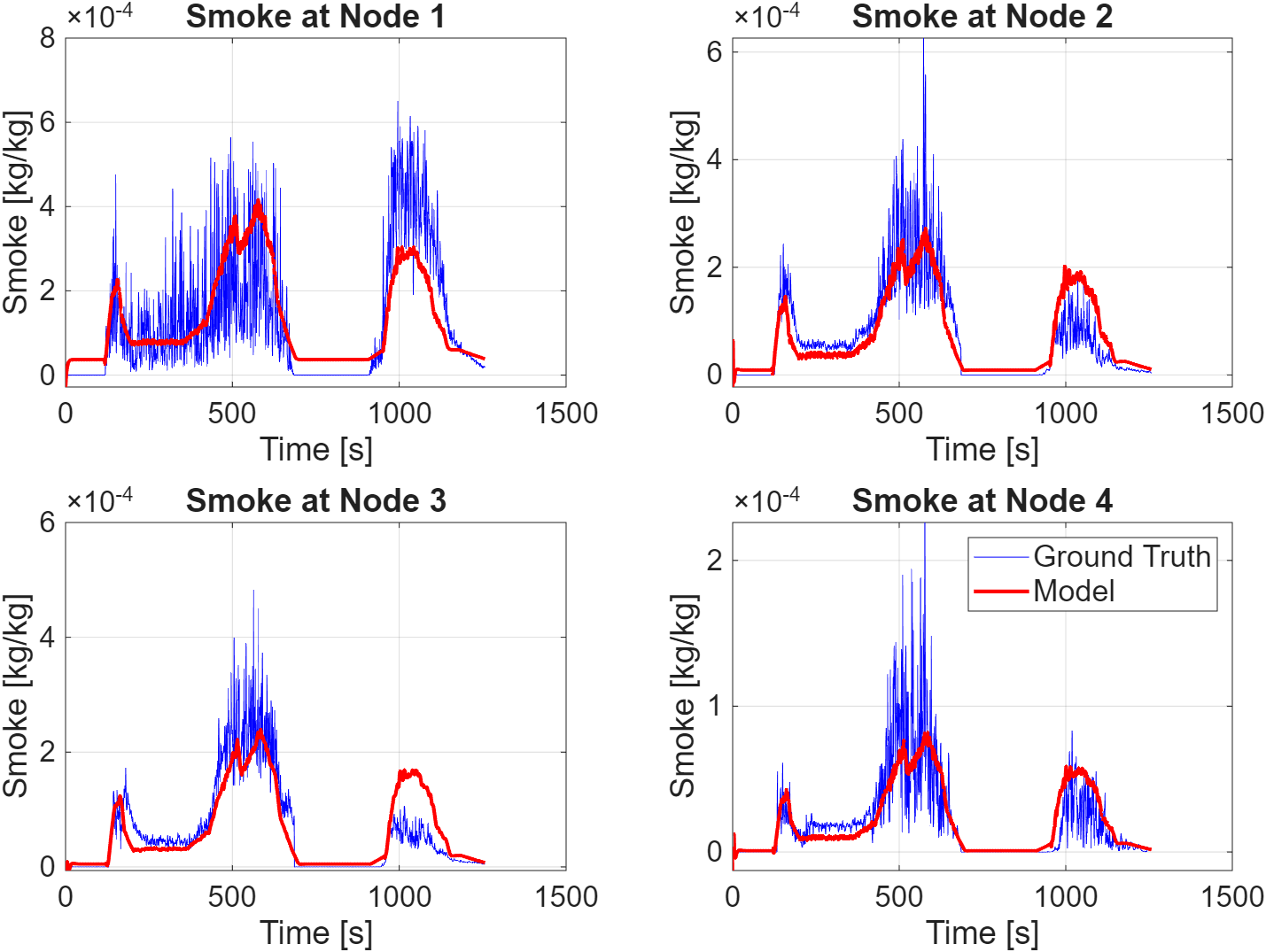}
    \caption{Comparison of the dynamical smoke propagation model and ground truth, for training dataset}
    \label{Train_smoke}
\end{figure}

\begin{figure}[h!]
    \centering
\includegraphics[width=0.7\textwidth]{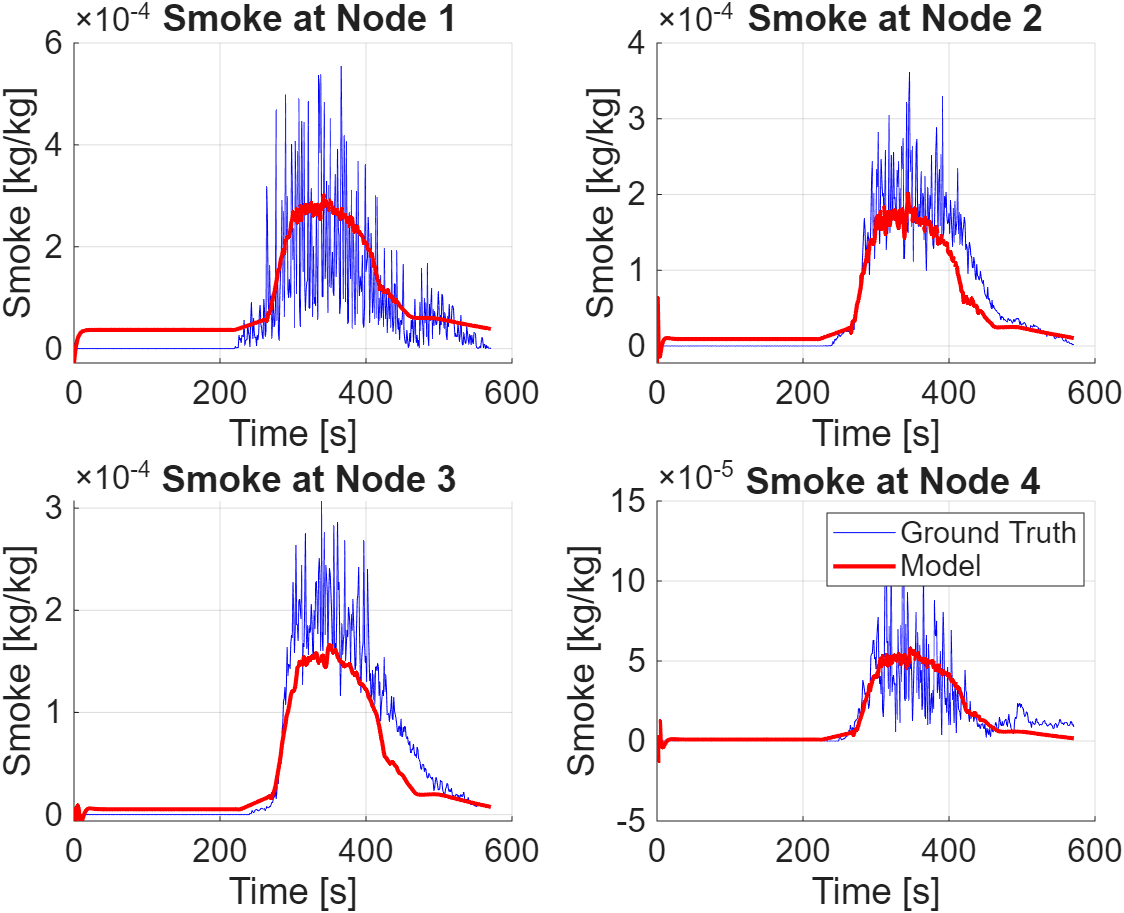}
    \caption{\centering Comparison of the dynamical smoke propagation model and ground truth obtained at the four nodes for testing dataset}
    \label{Testing_smoke}
\end{figure}

Concatenated data from the LIB TR simulation of 60 Ah and 243 Ah LIB under an airflow velocity of 2.0 m/s were used to train the state-space models using the System Identification Toolbox. 
The SSMs for temperature at the selected nodes are presented in Fig. \ref{Train}.
They have a fitting performance of 42.9\%, 68.1\%, 64.9\%, and 44.2\%, respectively.
Testing these models on a new airflow velocity data set of 1.0 m/s yielded reasonable accuracy for the nodes, as indicated in Fig. \ref{Testing}.
Similarly, the SSMs for smoke at the selected nodes are presented in Fig. \ref{Train_smoke}, and present a fitting performance of 29.8\%, 42.9\%, 44.8\%, and 36.8\%, respectively.
Testing these models on a new data set yielded reasonable accuracy for the nodes, as indicated in Fig. \ref{Testing_smoke}.

The apparent discrepancy between moderate model fitting and good visual trend alignment when using data from LES can be attributed to high-frequency perturbations and inherent turbulence built into LES models. 
LES resolves large-scale turbulent structures and models smaller eddies stochastically, resulting in the loss of smoothness, unlike the more conventional Reynold's-Averaged Navier-Stokes turbulence models that incorporate averaging. 
Therefore, in this instance, the visual agreement demonstrated by testing the models indicates the model's sound capability to reproduce dominant physical trends in the system.
The low numerical fit arises from the stochastic and highly unresolved nature of LES, which does not necessarily indicate the modeling error.

\subsection{Model uncertainty analysis }

\begin{figure}[h!]
    \centering
    \includegraphics[width=0.5\textwidth]{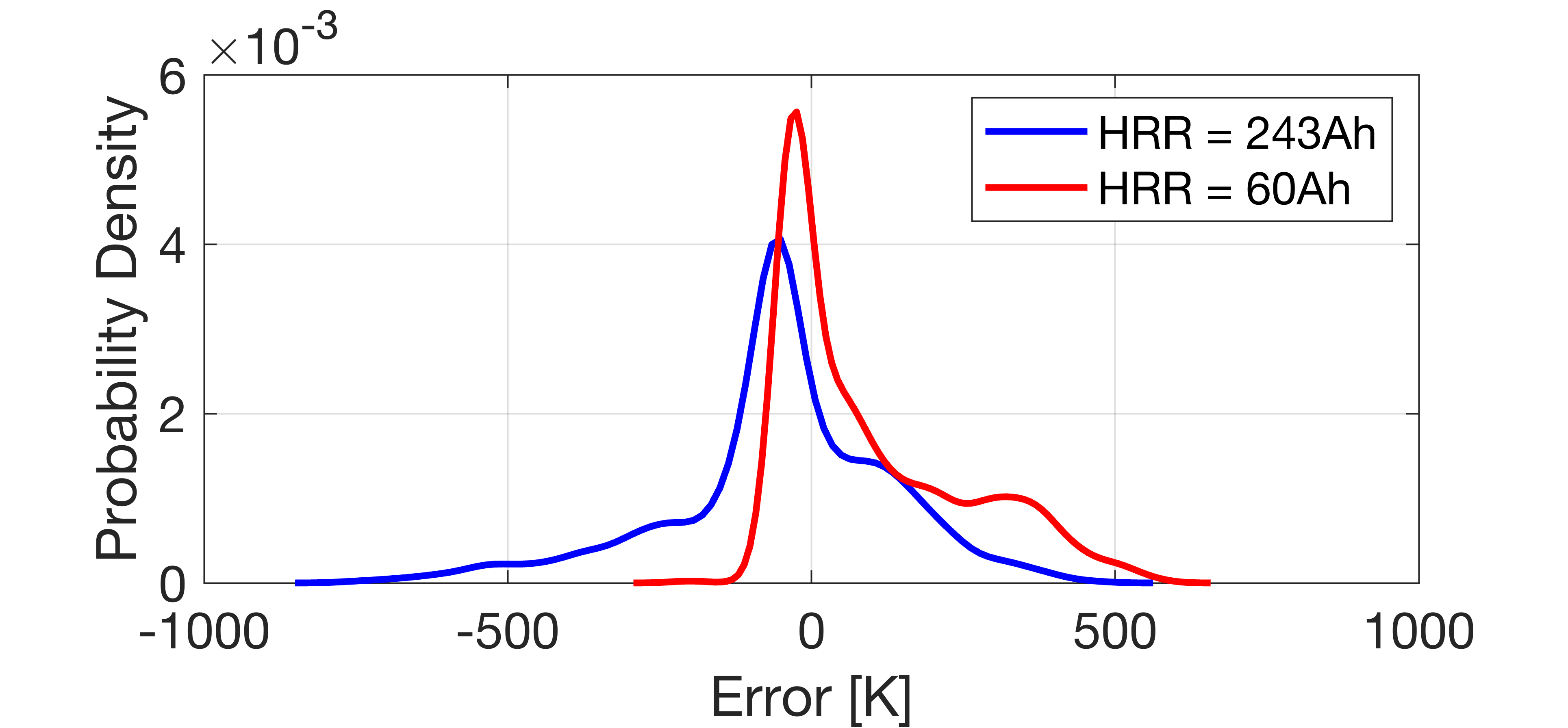}
    \caption{\centering Comparison of probability density estimate of the dynamical temperature model error with $HRR = 243Ah$ and $HRR = 60Ah$ at node 2}
    \label{UQT2}
\end{figure}

\begin{figure}[h!]
    \centering
    \includegraphics[width=0.5\textwidth]{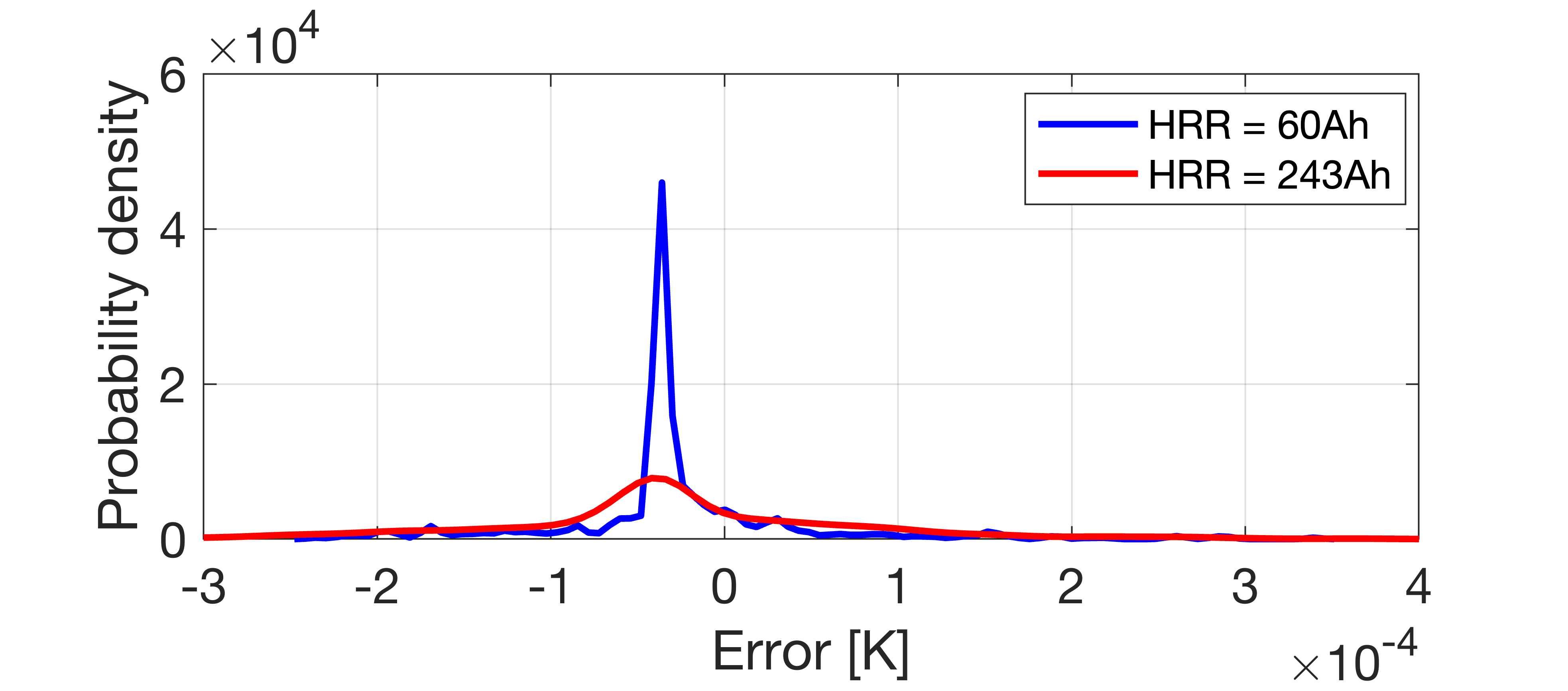}
    \caption{Comparison of probability density estimate of the dynamical smoke concentration model error with $HRR = 243Ah$ and $HRR = 60Ah$ at node 2. $T_{amb} = 25K$ }
    \label{UQS2}
\end{figure}

Probability density estimates (PDE) were used to assess the model uncertainty for selected nodes.
The residuals of the model were estimated using kernel density estimation.
For example, at node 2, the SSM for 60 Ah demonstrates a narrower and higher peak in its density curve, indicating a low uncertainty and greater confidence in its prediction.
However, the 243 Ah model indicates more sensitivity and variability. 
However, both curves are centered on the 0 K error, indicating that the models are of acceptable accuracy, as indicated in Fig. \ref{UQT2}.
Similar trends are observed for smoke concentration as indicated in Fig. \ref{UQS2}.
This exercise was performed for all other nodes and yielded similar results.

\begin{figure}[h!]
    \centering    \includegraphics[width=0.5\textwidth]{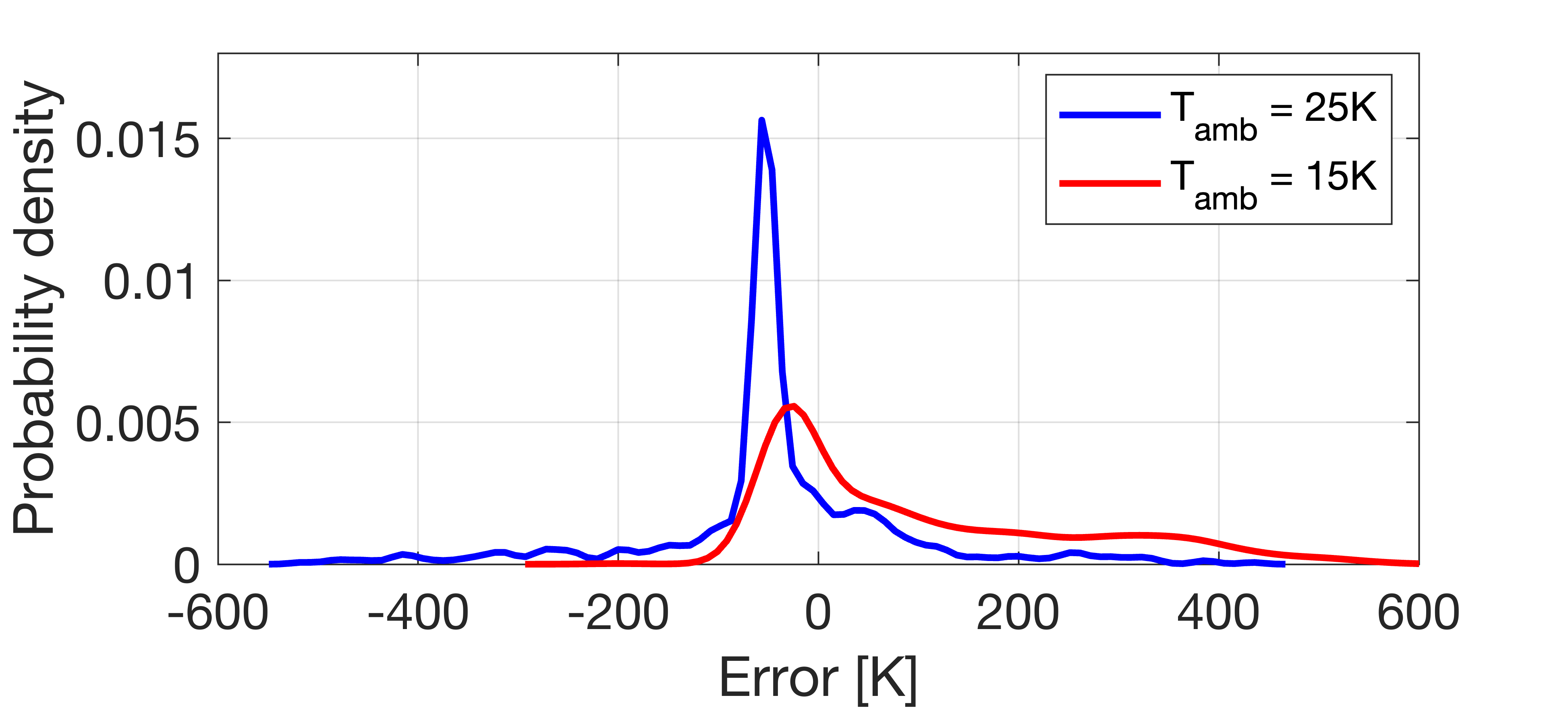}
    \caption{Comparison of the probability density estimate of 
    the dynamical temperature model error with $T_{amb} = 25K$ and $T_{amb} = 15K$ at node 2. $HRR = 60Ah$.}
    \label{UQT3}
\end{figure}

Furthermore, the influence of ambient temperature of \SI{15}{\celsius} and \SI{25}{\celsius} on model uncertainty was assessed at node 2 based on HRR of 60 Ah. 
The distributions derived from the kernel density estimation indicate that the model prediction error is centered around 0 K, indicating a generally reasonable accuracy.
However, the ambient temperature of \SI{25}{\celsius} indicates less variance and greater confidence in its estimates compared to the broader curve of \SI{15}{\celsius} as shown in Fig. \ref{UQT3}.
This suggests that lower ambient temperature introduces more uncertainty in model prediction, possibly from nonlinearity in heat transfer or LIB behavior at lower temperatures.

\section{Discussions \& Future Work} \vspace{2 mm}
\label{Future}

 Mines experience dynamical changes in their ventilation air flows due to variation in ambient atmospheric pressure, temperature, and humidity. This can be exacerbated by anomalous events, such as thermal runaway of a large-format battery. 
 In this work, a reduced order modeling approach has been presented that captures the changes in thermal and smoke dynamics after a battery fire in an underground mine tunnel. Specifically, a state-space modeling technique has been employed, which has been identified and tested using high-fidelity numerical simulation data obtained using Fire Dynamics Simulator software. 
The state-space model demonstrated strong visual alignment with the testing data, successfully capturing dominant transient trends.

High-fidelity models were developed for this research using the large eddy simulation (LES) framework. 
This approach resolves the thermo-fluid parameters on the grid larger than the selected cell size and models the ones smaller than it. 
Transient-state temperature and smoke exhibit rapid fluctuations in their magnitude due to sudden changes in geometry around the nodes. 
FDS also uses a structured grid with rectilinear elements which could have also contributed to the fluctuations.
Refining meshes further could have possibility have alleviated the fluctuations somewhat, but that would have come at a rapid increase in computational cost. 
Moderate numerical fitting scores (29.8–68.1\%) are attributed to high-frequency turbulence in LES-based training data. 
Furthermore, probability density estimates of model residuals centered around zero, which indicates no systematic bias.
The reduced order model framework provides a computationally efficient and sufficiently accurate method for tracking critical TR hazards, offering a practical tool for safety protocol development.

The work presented here could be extended to a much larger and more complex layout of underground mining operations. 
These form a complex network of connected tunnels and support structures. 
As the mines grow larger and deeper, geothermal gradients and autocompression of air add more heat to the ventilation air. 
Reduced-order models could be a powerful numerical tool for the mine operators in these cases, since developing high-fidelity three-dimensional models is computationally prohibitive. 
Changing any parameters in the boundary conditions will require the development of entirely new models. 
ROMs, on the other hand, can be rapidly developed for a suite of operating conditions. 
Finally, an ensemble of high-fidelity CFD models and ROMs provides a high degree of confidence as laboratory-scale testing of the numerical framework is planned. 
As indicated earlier, large-scale thermal runway testing is extremely hazardous and expensive. 
Efforts are underway to develop robust estimation models to mimic the post combustion conditions. 



\section*{Funding Source}
Thomas V. and Jean C. Falkie Mining Engineering Faculty Fellowship awarded to Dr. Ashish R. Kumar supported a part of this research.  


\newpage
\begingroup
\small
\setlength{\bibsep}{0pt}
\singlespacing
\bibliographystyle{ieeetr}
\bibliography{ref}
\endgroup

\end{document}